\documentclass[a4paper]{jpconf}
\usepackage{graphicx}
\usepackage{amsfonts}
\usepackage{amsmath}
\begin{document}
\title{Mutually Unbiased Bases and Semi-definite Programming}

\author{Stephen Brierley$^{1}$ and Stefan Weigert$^{2}$}

\address{$^1$ QuIC, Ecole Polytechnique, Université Libre de Bruxelles, CP 165, 1050 Brussels, Belgium} 
\address{$^2$ Department of Mathematics, University of York, York YO10 5DD, UK}

\ead{steve.brierley@ulb.ac.be, slow500@york.ac.uk}

\begin{abstract}
A complex Hilbert space of dimension six supports at least \emph{three} but not more than \emph{seven} mutually unbiased bases. Two computer-aided \emph{analytical} methods to tighten these bounds are reviewed, based on a discretization of parameter space and on Gr\"obner bases. A third algorithmic approach is presented: the non-existence of more than three mutually unbiased bases in composite dimensions can be decided by a \emph{global} optimization method known as semidefinite programming. The method is used to confirm that the \emph{spectral matrix} cannot be part of a complete set of seven mutually unbiased bases in dimension six. 
\end{abstract}

\begin{flushright}
\end{flushright}

\section{Introduction}
Two orthonormal bases $\mathcal{B}^{0}=\{|\psi _{j}^{0}\rangle ,j=1\ldots d\}
$ and $\mathcal{B}^{1}=\{|\psi _{j}^{1}\rangle ,j=1\ldots d\}$ of $\mathbb{C}%
^{d}, d \in \mathbb{N}$, are mutually unbiased (MU) if the modulus of the inner product of any vectors not in the same basis is constant, 
\begin{equation} 
|\langle \psi _{i}^{0}|\psi _{j}^{1}\rangle |=\frac{1}{\sqrt{d}},\text{ \
for all \ }i,j=1\ldots d.  \label{MU conditions}
\end{equation}
In this paper, we will examine three computer-aided analytical methods which could be used to prove the conjecture that no more than three MU bases exist in the space $\mathbb{C}^{6}$ \cite{zauner99}. This open problem concerning the geometry of the state space of a low-dimensional quantum system has attracted considerable interest in recent years; see \cite{durt+10} for a thorough review of the properties and uses of MU bases in complex Hilbert spaces of dimension $d$.

First, we review and illustrate an approach which is based on \emph{discretising the  parameter space} when searching for vectors that constitute MU bases. This simplification results in only a \emph{finite}---but possibly large---number of states which must be checked. The results obtained will be exact once rigorous error bounds have been established \cite{jaming+09}. Second, we show that the constraints defining an MU constellation can be expressed in terms of coupled polynomial equations. Thus, constructing a \emph{Gr\"{o}bner basis} may show that they have no solution. Finally, we propose a new approach by writing the problem in a form suitable to use \emph{semidefinite programming}. This allows one to algorithmically identify the (non-) existence of solutions by means of global optimization techniques. 

We will illustrate each of the proposed methods by reproducing known results for MU bases in low dimensions \cite{brierley+09b}. For the sake of simplicity, each of the algorithms will be used to prove the non-existence of four MU bases in dimension two. All three techniques have yet to be applied successfully to the problem in dimension six: the \emph{computational} resources required to implement them were beyond those available.

Throughout this paper, we will use the concept of \emph{MU constellations} 
\cite{brierley+08}: the MU constellation $\{ d-1,\lambda ,\mu ,\nu \}_{d}$ with $0\leq \lambda ,\mu ,\nu \leq d-1$, for example, consists of four sets of orthonormal vectors containing $d-1,\lambda ,\mu$, and $\nu$ vectors, respectively, with vectors taken from different sets having the required overlap $1/\sqrt{d}$ (cf. Eqs. (\ref{MU conditions})). It is sufficient to list at most $(d-1)$ vectors in each set since a suitable $d^{\rm{th}}$ vector can be constructed from them.

The non-existence of any MU constellation smaller than $\{5,5,5,5\}_{6} \equiv \{5^4\}_6$ proves the non-existence of four (or more) MU bases in dimension six. The smallest number of vectors sufficient to parameterize a MU constellation $\{d-1,\lambda ,\mu ,\nu \}_{d}$ is given by $s=\lambda +\mu +\nu -1$ (the underlying equivalence relations between MU constellations can be found in Appendix A of \cite{brierley+08}). The number $s$ effectively determines the number of variables and equations needed to define a MU constellation and, \emph{a fortiori}, the computational resources required to run any algorithm solving the problem. For example, the MU constellation $\{5,3,3,3\}_{6} \equiv \{5,3^3\}_6$ contains $s=8$ free vectors depending on 40 real parameters, whereas the full set of seven MU bases in dimension six involves $s=29$ vectors depending on 145 real parameters. In order to reduce the problem as much as possible, it seems imperative to work with smallest possible constellations such as $\{5,3^3\}_6$ or $\{5^2,4,1\}_6$ which have \emph{not} been found by the extensive numerical searches presented in \cite{brierley+08}. 

\section{Discretizing the parameter space}

One of the few rigorous results about MU bases in $\mathbb{C}^6$ is due to Jaming et al. \cite{jaming+09}. They have shown that a particular class of matrices, the so-called \emph{Fourier family} \cite{haagerup96}, cannot be a member of a quadruple of MU bases. The Fourier family consists of $(6 \times 6)$ matrices $F(x_{1},x_{2})$ depending on two continuous parameters $x_1$ and $x_2$; for $x_1=x_2=0$, the general expression reduces to the standard Fourier matrix $F_{jk} = \omega^{jk}/\sqrt{6}, j,k = 0\ldots 5$, where $\omega = \exp (2\pi i/6)$. We now describe the idea of this approach which, as outlined in \cite{matolci+09}, could lead to a no-go theorem in composite dimensions such as $d=6$.

Let us assume that there exists a MU constellation $C$ of the form $\{d-1,\lambda ,\mu  ,\nu \}_{d}$; we will attempt to find a contradiction. The constellation $C$ can be parameterised by $p_{d}$ phases $\mathbf{\alpha } \equiv (\alpha _{1},\ldots ,\alpha _{p_{d}})^{T}$, according to Eq. (8) of \cite{brierley+08}. The first step is to \emph{approximate} this MU
constellation by another set of vectors with components being restricted to $N^{\rm{th}}$ roots of unity only. Such an  approximation is easily achieved by dividing the interval $[0,2\pi )$ into $N$ non-overlapping intervals $I_{j}=[(2j-1)\pi/N,(2j+1)\pi/N)\;  (\!\!\!\mod 2\pi)$, with  $j=1,\ldots ,N$. If the phase $\alpha _{k}$ lies in the interval $I_{j},$ we approximate it by the mid-point of $I_{j},$ $\alpha _{k}\rightarrow \widetilde{\alpha }_{k,j}\equiv 2\pi j_k/N.$ The mapping $\mathbf{\alpha }\rightarrow \widetilde{\mathbf{\alpha }}$ thus sends the vectors of a given MU constellation $C$ to a different collection of vectors (denoted by $\widetilde{C}$) which do not necessarily satisfy the MU conditions (\ref{MU conditions}). Effectively, the $p_d$-dimensional continuous parameter space $\mathbf{\alpha}$ has been replaced by a grid consisting of a \emph{finite} number of 
$N^{p_d}$ states. The accuracy of the approximation of the MU constellation $C$ by $\widetilde{C}$ is determined by the value of $N$, and additional flexibility results from partitioning the range of each variable $\alpha_k$ individually.  

The key step forward made by Jaming et al. \cite{jaming+09} has been to establish rigorous bounds on the errors of the scalar products introduced by the discretization. These bounds allow one to conclude that no MU constellation exists if none can be found for a sufficiently accurate discrete approximation. Thus, the search for a contradiction has effectively been reduced to checking a \emph{finite} set of constellations which can be done exhaustively, in principle. If $N$ is small, the search is easy to perform; however, the error bounds are not tight for small values of $N$. Finer partitions are necessary which, in turn, come at a computational cost since the number of grid points is proportional to $N^{p_d}$. A hierarchical refinement of the procedure may be used to reduce the amount of computational resources \cite{matolci+09}.

\subsection{Dimension two}
It is not difficult to illustrate this algorithm by searching for the MU constellation $\{1^4\}_2$, i.e. a set of four MU bases in the space $\mathbb{C}^2$---which is known \emph{not} to exist. If the constellation $\{1^4\}_2$ existed, the four vectors defining it could be written in the form
\begin{equation} \label{fourvectors}
\left( 
\begin{array}{c}
1 \\ 
0%
\end{array}%
\right) ,\frac{1}{\sqrt{2}}\left( 
\begin{array}{c}
1 \\ 
1%
\end{array}%
\right) ,\frac{1}{\sqrt{2}}\left( 
\begin{array}{c}
1 \\ 
e^{i\alpha}%
\end{array}%
\right) ,\frac{1}{\sqrt{2}}\left( 
\begin{array}{c}
1 \\ 
e^{i\beta}%
\end{array}%
\right) \, ,
\end{equation}
depending on $p_2 = 2$ real parameters $\alpha (\equiv \alpha_1)$ and $\beta (\equiv \alpha_2)$. The conditions resulting from Eqs. (\ref{MU conditions}) lead to three constraints,
\begin{equation} \label{MUconditionsexample1}
\left| \cos \frac{\alpha}{2} \right| 
     = \left| \cos \frac{\beta}{2} \right| 
     = \left| \cos \frac{\alpha - \beta}{2} \right| 
     = \frac{1}{\sqrt{2}}
\end{equation}
and it is obvious that they have no solution. 

To see the discretization procedure in action, we now show the inconsistency of Eqs. 
(\ref{MUconditionsexample1}) by checking only a finite number of cases. Assume that there are values of $\alpha$ and $\beta$ which give rise to a solution of Eqs. (\ref{MUconditionsexample1}), and write them as $\alpha = \alpha_j + \Delta \alpha_j$ and  $\beta = \beta_{j^\prime} + \Delta \beta_{j^\prime}$, where $\alpha_j$ is the  $N^{\rm{th}}$ root of unity closest to $\alpha$ and differing by $\Delta \alpha_j$ from it, etc. The maximum distance between the actual value of $\alpha$ and an $N^{\rm{th}}$ root is $\pi/N$. Using Taylor's theorem for a differentiable function $f(x)$, 
\begin{equation} \label{Taylor}
f(x+h) = f(x) + h f^\prime (x+\theta (x) h) \, , \quad \mbox{for some } \theta (x) \in (0,1) \, , 
\end{equation}
the condition $| \cos(\alpha/2)| = 1/\sqrt{2}$ implies that
\begin{equation} \label{MUconditionsexample1approx}
\frac{1}{\sqrt{2}} - \frac{\pi}{2N} 
     \leq \left| \cos \frac{\alpha_j}{2} \right|
     \leq \frac{1}{\sqrt{2}} + \frac{\pi}{2N} \, ,
\end{equation}
where the triangle inequality 
\begin{equation} \label{triangle}
|a_1| - |a_2| \leq |a_1 + a_2| \leq |a_1| + |a_2| \, , 
\end{equation}
and the relations $|\Delta \alpha_j| \leq \pi/N$ have been used. The other two constraints lead to inequalities which are obtained by substituting either $\beta_{j^\prime}$ or $\alpha_j-\beta_{j^\prime}$  for $\alpha_j$ in (\ref{MUconditionsexample1approx}). Thus, one needs to check $N^2$ inequalities corresponding to the indices $j,j^\prime = 1 \ldots N$. For increasing values of $N$, the numbers $\alpha_j, \beta_{j^\prime}$, \emph{and}   $\alpha_j-\beta_{j^\prime}$ must approach odd multiples of $\pi/2$ which is impossible. Thus, for a sufficiently large \emph{finite} value of $N$, the inequalities have no solutions in terms of $N^{\rm{th}}$ roots of unity and in some intervals centered around them---thus covering the \emph{entire} parameter space. 

\subsection{Dimension six}
Jaming et al. have used this method to exclude the entire Fourier family, $F(x_1,x_2)$,
from a complete set of MU bases \cite{jaming+09}. This was achieved by discretising the
fundamental region of the Fourier family using $N=180$ and two other
complete bases using $N^{\prime }=19$. The restriction to consider sets of
the form $\{I,F(x_{1},x_{2}),\mathcal{B}^{2},\mathcal{B}^{3}\}$ with two orthonormal bases $\mathcal{B}^{2}$ and $\mathcal{B}^{3}$, reduces the
computational complexity considerably but it is not dictated by an inherent limitation of the method. Interestingly, the non-existence of a finite projective plane was shown by an exhaustive search \cite{lam+89}. Since the existence of a complete set of MU bases shares some properties with the existence of finite projective planes \cite{saniga+04}, discretization appears
a promising avenue.

\section{Using Gr\"{o}bner bases \label{GB global sol}}

The second method we review relies on a technique to algorithmically search for solutions of coupled polynomial equations. The approach is based on the construction of a \emph{Gr\"obner basis} \cite{buchberger65} for these equations, a tool developed in commutative algebra. It has been applied successfully in \cite{grassl05} to show that no two MU bases of a particular form can be supplemented by a third basis plus a single vector which is MU to those of the three bases. Further rigorous results excluding many candidate bases from a complete set of MU bases have been reported in \cite{brierley+09}. We first formulate the approach for arbitrary dimensions $d$, then work through an explicit example in the space $\mathbb{C}^2$, and finally discuss the prospects of applying this approach in dimension six.  

Any MU constellation of type $\{d-1,\lambda ,\mu ,\nu \}_{d}$ will correspond to a point on a $s(d-1)$-dimensional  hypertorus, with $s=\lambda +\mu +\nu - 1$ \cite{brierley+08}. In a parameterization analogous to the one given in (\ref{fourvectors}), only phase factors of modulus one each depending on a single real parameter will occur. Upon expressing each phase factor as $e^{i\alpha_{j}} = x_{j}+iy_{j}$, supplemented by the conditions $x_{j}^{2}+y_{j}^{2}=1$, we are led to a parameterization which requires $2s(d-1)$ real variables ${\bf x} \equiv (x_1, x_2, \ldots, y_{2s(d-1)}) \in \mathbb{R}^{2s(d-1)}$. 

The conditions for vectors, orthonormal in sets with $d-1,\lambda ,\mu$ and $\nu$ 
elements, to be MU now turn into $N$ multivariate polynomial equations,
\begin{equation} \label{polynomials}
p_{j}({\bf x}) = 0 \, , \qquad j=1\ldots N \, , 
\end{equation}
with
\begin{equation} \label{numberofequations}
N = \frac{1}{2}(s+1)(s-1)\,+\frac{1}{2}(\lambda ^{2}+\mu ^{2}+\nu^{2})+s(d-1) \, .
\end{equation}
Any solution of the Eqs. (\ref{polynomials}) gives rise to a MU constellation of type $\{d-1,\lambda ,\mu ,\nu \}_{d}$. Geometrically speaking, we wish to describe the variety 
\begin{equation} \label{variety}
V\equiv \{\mathbf{x}\in \mathbb{R}^{2s(d-1)}:p_{1}(\mathbf{x})=0,\ldots ,p_{N}(\mathbf{x}) =0\} \, ,
\end{equation}
which is a subset of the space $\mathbb{R}^{2s(d-1)}$. No MU constellation of the form $\{d-1,\lambda ,\mu ,\nu \}_{d}$ exists if and only if the associated variety is empty, $V=\emptyset$. The set $V$ has an algebraical description in terms of an ideal $I=\langle p_{1},\ldots ,p_{N}\rangle$ which is generated by the polynomials $p_{1}(\mathbf{x})$ to $p_{N}(\mathbf{x})$. This ideal $I$ consists of all linear combinations $a$ of the polynomials $p_{j}(\mathbf{x}) $ with coefficients $r_{j} $ polynomial in the variables $\mathbf{x}$, 
\begin{equation} \label{ideal}
a=\sum_{j=1}^{S}r_{j}(\mathbf{x})p_{j}(\mathbf{x}) \, .
\end{equation}
It is important to note that the ideal corresponds to the variety over the algebraic closure of the coefficient field, here the \emph{complex} numbers.

Having re-cast the problem of identifying constellations of MU vectors in terms of ideals, it becomes possible to apply methods from commutative algebraic geometry. The equations (\ref{polynomials}) have no (real or complex-valued) solutions if the polynomial $1$ is contained in the ideal $\langle p_{1},\ldots ,p_{N}\rangle$. The construction of a Gr\"{o}bner basis $G$ of the ideal $I$ would allow us to check this property since $G$ would be given by the set $G=\{1\}$. The converse is not necessarily true: the equations defining a MU constellation may have no solutions over the real numbers but the ideal is non-empty due to the existence of complex solutions. The equation $x^{2}+1=0$, for example, has no \emph{real} solutions but the ideal $I=\langle x^{2}+1\rangle $ is not generated by $\{1\}$. In a fortunate situation, the variety is found to be empty over the complex numbers which would then constitute a proof that a complete set of MU bases does not exist in dimension six.

\subsection{Dimension two}

We now confirm the validity of the algorithm based on Gr\"obner bases by proving (again) that no four MU bases exist in dimension two. First, we write down the system of coupled polynomial equations, the solutions to which would define the MU
constellation $\{1^{4}\}_{2}$.  \emph{Four} real variables $\{x_{1},x_{2},y_{1},y_{2}\}$ parameterize the candidate vectors shown in Eq. (\ref{fourvectors}),
\begin{equation} \label{fourvectors2}
\left( 
\begin{array}{c}
1 \\ 
0%
\end{array}%
\right) ,\frac{1}{\sqrt{2}}\left( 
\begin{array}{c}
1 \\ 
1%
\end{array}%
\right) ,\frac{1}{\sqrt{2}}\left( 
\begin{array}{c}
1 \\ 
x_{1}+iy_{1}%
\end{array}%
\right) ,\frac{1}{\sqrt{2}}\left( 
\begin{array}{c}
1 \\ 
x_{2}+iy_{2}%
\end{array}%
\right) \, ,
\end{equation}
supplemented by the constraints
\begin{eqnarray} \label{modulus1}
p_{1}(\mathbf{x}) &\equiv &x_{1}^{2}+y_{1}^{2}-1=0 \, , \notag \\
p_{2}(\mathbf{x}) &\equiv &x_{2}^{2}+y_{2}^{2}-1=0 \, .
\end{eqnarray}
The constraints defining a MU constellation $\{1^{4}\}_2$ read explicitly
\begin{eqnarray}
p_{3}(\mathbf{x}) &\equiv &(1+x_{1})^{2}+y_{1}^{2}-2=0 \, , \notag \\
p_{4}(\mathbf{x}) &\equiv &(1+x_{2})^{2}+y_{2}^{2}-2=0 \, ,  
\label{MUconditionsexample2}\\
p_{5}(\mathbf{x}) &\equiv
&(1+x_{1}x_{2}+y_{1}y_{2})^{2}+(x_{1}y_{2}-x_{2}y_{1})^{2}-2=0 \, .  \notag
\end{eqnarray}
It is straightforward to derive that these equations have no solution, a fact which, for the purpose of illustration, we will confirm by showing that they have $G=\{1\}$ as a Gr\"obner basis.

We have constructed the Gr\"{o}bner basis for Eqs. (\ref{modulus1},\ref{MUconditionsexample2}) using a package called FGb \cite{faugere99} implemented in Maple \cite{Maple}. A standard desktop PC outputs the desired Gr\"{o}bner basis in 0.016 seconds, and it is indeed equal to all polynomials over the complex numbers,
\begin{equation}
\langle p_{1}(\mathbf{x}),\ldots ,p_{5}(\mathbf{x})\rangle =\langle 1\rangle
.
\end{equation}
A direct confirmation of this result follows from writing down the coefficient polynomials used to construct the Gr\"{o}bner basis, namely
\begin{eqnarray}
r_{1}(\mathbf{x}) &\equiv &-\frac{1}{2}\left( y_{1}y_{2}+x_{1}+2\right) 
\notag \, , \\
r_{2}(\mathbf{x}) &\equiv &-\frac{1}{2}\left(
x_{1}^{2}y_{2}+y_{1}^{2}y_{2}-x_{1}y_{2}+x_{2}y_{1}+2y_{1}\right) y_{1} \, , 
\notag \\
r_{3}(\mathbf{x}) &\equiv &\frac{1}{2}x_{1} \, , \\
r_{4}(\mathbf{x}) &\equiv &-\frac{1}{2}\left( x_{1}y_{2}-x_{2}y_{1}\right)
y_{1} \, ,  \notag \\
r_{5}(\mathbf{x}) &\equiv &\frac{1}{2}y_{1}y_{2} \, .  \notag
\end{eqnarray}
It is not difficult to verify that the following relation holds,
\begin{equation} \label{explicitform}
\sum_{j=1}^{5}r_{j}(\mathbf{x})p_{j}(\mathbf{x})=1 \,,
\end{equation}
which establishes that $G \equiv \{1\}$. It  follows that the variety $V=\{\mathbf{x}\in  \mathbb{R}^{4}: p_{1}(\mathbf{x})=0, \ldots ,p_{5}(\mathbf{x})=0\}$ is empty implying that no four MU bases exist in the space $\mathbb{C}^2$.  

\subsection{Dimension six}

Unfortunately, 16GB of memory have not been sufficient to decide if the Gr\"{o}bner
bases corresponding to the equations generated by the constellations $
\{5,3^3\}_6$ or $\{5^2,4,1\}_6$ contain the element $1$. The
computations run out of memory \emph{before} the algorithm terminates since the number of variables and equations is considerable, even for these smallest possible interesting constellations in dimensions six. For example, the constellation $\{5^2,4,1\}_6$ requires the construction of a Gr\"{o}bner basis of an ideal generated by 61 equations of degree 4 in 90 real variables.

\section{Using semidefinite programming}

Semidefinite programming \cite{vandenberghe+96} is a powerful tool to obtain rigorous results aided by a computer which has been used in areas such as control theory and combinatorial optimization \cite{vandenberghe+99}. Recent years have seen first  applications in quantum information theory to solve problems such as deciding whether a given mixed state $\rho $ is entangled or not 
\cite{brandao+04} - \cite{doherty+04}. The success of semidefinite
programming stems from the fact that the type of problems studied are \emph{efficiently} solvable by a computer. In addition, there
is a duality theorem which provides a \emph{certificate} allowing one to directly check the result. For example, when applied to the separability problem, a
semidefinite program decides whether a given mixed state is entangled---if it
is, it outputs an \emph{entanglement witness} \cite{horodecki+09}, that is, a hyperplane which separates the entangled state at hand from the set of all separable states.

Another interesting application of semidefinite programming to problems about
finite dimensional Hilbert spaces is to the \emph{compatibility problem} for tripartite quantum systems. Here one asks if there exists a single state of the entire system given the states of all (proper) reduced states. Hall \cite{hall06} has cast the compatibility problem in the form a semidefinite program and used it to disprove a conjecture of Butterley et al. \cite{butterley+06}. In this case, the solution to the dual problem resulted in a certificate, called an \emph{incompatibility witness}, which proves that a particular set of reduced states are not compatible with any multipartite state.

A \emph{semidefinite program} (SDP) is an algorithm for solving an
optimization problem of the form
\begin{eqnarray}
&&\text{minimise }\mathbf{c}^{T}\mathbf{x} \notag \\
&&\text{subject to }F(\mathbf{x})\geq 0 \label{sdpgeneral}
\end{eqnarray}
where $\mathbf{c}$ is a fixed vector,  and the \emph{decision variables} $\mathbf{x}$ are constrained by the requirement that the matrix $F(\mathbf{x})\equiv
x_{1}F_{1}+\cdots +x_{n}F_{n}-F_0$ be positive semidefinite, with symmetric $n\times n$ matrices $F_0,F_{1},\ldots ,F_{n}$ \cite{vandenberghe+96}. This is a \emph{convex} optimization problem. A linear program, where the constrains have the form $F(\mathbf{x} )= \rm{diag}(A\mathbf{x-b),}$ is an example of a SDP with important applications in economics \cite{mathiesen85} or finding optimal network flows \cite{bertsimas+03}. 

A SDP can also be used to solve \emph{non-linear} problems as long as they are convex. For example, the problem
\begin{eqnarray}
&&\text{minimise }\frac{(\mathbf{c}^{T}\mathbf{x)}^{2}}{\mathbf{d}^{T}%
\mathbf{x}} \notag \\
&&\text{subject to }A\mathbf{x}\geq \mathbf{b} \label{sdp1}
\end{eqnarray}
can be re-written as a SDP \cite{vandenberghe+96},
\begin{eqnarray}
&&\text{minimise }t \notag \\
&&\text{subject to }\left( 
\begin{array}{ccc}
\rm{diag}(A\mathbf{x-b}) & 0 & 0 \\ 
0 & t & \mathbf{c}^{T}\mathbf{x} \\ 
0 & \mathbf{c}^{T}\mathbf{x} & \mathbf{d}^{T}\mathbf{x}%
\end{array}%
\right) \geq 0 \, . \label{sdp2}
\end{eqnarray}

It is possible to cast the existence problem of MU constellations as a semidefinite program \cite{brierley09}.  The equations $p_{j}(\mathbf{x} )=0$ defining MU constellations are, unfortunately, not convex since they involve fourth-order polynomials. However, all is not lost: it is, possible to extend the applicability of this method by using tools from \emph{non-convex optimization} \cite{kojima+00, shorN87}. Lasserre has shown that, upon relaxing the non-convex constraints, one can define a \emph{hierachy} of semidefinite programs representing ever better approximations to the original one \cite{lasserre01}, at the cost of introducing additional decision variables. At each level, this \emph{method of relaxations} either rules out the existence of a MU constellation or is inconclusive requiring an additional step of relaxation. Each iteration inevitably leads to a computationally more difficult problem but the remarkable work of Lasserre \cite{lasserre01} ensures that after a finite number of iterations the \emph{exact} solution will have been obtained. In other words, the hierarchy is \emph{asymptotically complete}. It is this type of approach which has been applied successfully to problems from quantum information \cite{eisert+04}.

Let us now show how to express the existence problem of a MU constellation  as an optimization problem. To do so, 
\begin{quote} 
\emph{pick any one of the polynomials from (\ref{polynomials}), say $p_{1}(\mathbf{x})$, and find the minimum value of $p^2_{1}(\mathbf{x})$ subject to the condition that the variables satisfy the remaining constraints, $p_{2}(\mathbf{x})=0,\ldots ,p_{N}(\mathbf{x})=0$.}
\end{quote}
Effectively, we seek to minimise the value of $p_{1}^2(\mathbf{x})$ in the solution space of the remaining polynomials. Lasserre's method of relaxations then allows one to find a lower bound $B_{L}(r)$ for the function $p_{1}^2(\mathbf{x})$,  where $r$ is the degree of the relaxation. If for some degree of relaxation, $r=2,3,\ldots$ one finds a \emph{positive} bound, 
$B_{L}(r)>0$, then \emph{no} MU constellation of the desired type exists. The dual formulation of the SDP would automatically provide a certificate of non-existence, offering an independent way of verifying the result. However, the original problem is transformed by the relaxation steps and so this ``witness of non-existence'' is less intuitively connected to the existence of an MU constellation than, for example, an entanglement witness.


Here is a sketch of the algorithm which is capable to determine whether a MU constellation $\{d-1,\lambda ,\mu .\nu \}_{d}$ exists:

\begin{enumerate}
\item write down the polynomial equations which define the MU constellation;

\item generate a SDP at the lowest possible level of relaxation ($r=2$);

\item solve the resulting SDP---if a positive global lower bound $B_{L}(r)$ is found then no MU constellation $\{d-1,\lambda ,\mu .\nu \}_{d}$ exists; otherwise, repeat Steps (ii) and (iii) at the next level of relaxation, $r:=r+1$.
\end{enumerate}
If no MU constellation exists, the algorithm is guaranteed to
find a positive lower bound. As $r$ increases, the global lower bounds $%
B_{L}(r)$ converge monotonically to the exact global minimum of the
function, $B_{L}^{opt}$. If the sought-after MU constellation does exist, the
algorithm will find an explicit parameterisation thereof, with a high level of
numerical accuracy.

\subsection{Dimension two}

To begin, we now spell out and run a SDP algorithm to show once more that that there are no more than three MU bases in dimension two. More explicitly, we will calculate a \emph{strictly positive} global lower bound $B_{L}(r)$ for the polynomial
\begin{equation}
p_{1}^{2}(\mathbf{x})=\left( x_{1}^{2}+y_{1}^{2}-1\right) ^{2} \, ,
\end{equation}
subject to the requirement that the remaining polynomials in Eqs. (\ref{modulus1}) and (\ref{MUconditionsexample2}) vanish (Step (i)). This leaves us with the following minimization problem,
\begin{eqnarray}
\min &&\left( x_{1}^{2}+y_{1}^{2}-1\right) ^{2}  \notag \\
\text{subject to } &&x_{2}^{2}+y_{2}^{2}-1=0  \notag \\
&&(1+x_{1})^{2}+y_{1}^{2}-2=0  \label{dim2 sdp} \\
&&(1+x_{2})^{2}+y_{2}^{2}-2=0  \notag \\
&&(1+x_{1}x_{2}+y_{1}y_{2})^{2}+(x_{1}y_{2}-x_{2}y_{1})^{2}-2=0\,   \notag
\end{eqnarray}
with four decision variables $\mathbf{x} = (x_1, \ldots, y_2)$. The MU constellation $\{1^{4}\}_{2}$ exists if and only if no global lower bound of $p_{1}^{2}(\mathbf{x})$ exceeds the value zero, or $B_{L}(r)\leq 0$ for all $r$. Any \emph{positive} bound implies the non-existence of a MU constellation $\{1^{4}\}_{2}$.

Using the Matlab package \texttt{gloptipoly3} \cite{gloptipoly3} which is
based on the theory presented in \cite{lasserre08}, we are able to convert the
problem (\ref{dim2 sdp}) into a semidefinite program, thus completing Step (ii). The resulting SDP has been solved using the \texttt{SeDuMi} MatLab package developed by Strum et al. \cite{SeDuMi}, corresponding to Step (iii). The results of the computations at three levels of relaxation, $r=2,3,4$, are presented in Table \ref{tableChap5}. We find that already the first level of relaxation, $r=2$, provides a positive lower bound for the polynomial $p_{1}^2(\mathbf{
x})$, $B_{L}(2)>0$. This result confirms that semidefinite programming is capable to positively identify non-existing MU constellations. 

The third and fourth columns of Table \ref{tableChap5} show how the size of the SDP grows as we increase the level of relaxation. The number of decision variables, $N_{d}$, jumps from 69 to 494. Similarly, the dimension of matrices defining the
semidefinite constraint $F(\mathbf{x})\geq 0$ is almost five times larger at 
$r=4$ than at the lowest level of relaxation, $r=2$. The algorithm proves to
be very efficient taking only 0.11 seconds on a desktop PC to convert the
original problem and solve the SDP for $r=2$. This time rises to 1.67 seconds
when $r=4$.\\

\begin{table}[th]
\begin{center}
\begin{tabular}{cccc}
\hline\hline
$r$ & $B_{L}$ & $N_{d}$ & $F$ \\ \hline
$2$ & $1.4038\times 10^{-8}$ & $69$ & $15\times 15$ \\ 
$3$ & $0.5359$ & $209$ & $35\times 35$ \\ 
$4$ & $0.5359$ & $494$ & $70\times 70$ \\ \hline\hline
\end{tabular}%
\end{center}
\caption[Lower bounds of the minimization problem in dimension two]{Lower bounds of the minimization problem defined in Eq. (\ref{dim2 sdp}), with levels of relaxation denoted by $r$, while $N_{d}$ and $F$ give the number of decision variables and the size of the semidefinite inequalities in the resulting SDP, respectively.}
\label{tableChap5}
\end{table}

The high level of numerical accuracy achieved by the optimization program 
\texttt{SeDuMi} allows us, in fact, to find an analytic expression for the lower
bound, namely $B_{L}^{opt}=(1-\sqrt{3})^{2}$. At the third and fourth levels of relaxation, the lower bound is optimal in that the function $p_{1}^2(\mathbf{x})$ reaches the value $B_{L}.$ It is
possible to output the parameter values $\mathbf{x}$ which achieve this
lower bound. There are two sets of vectors corresponding to the global minimum value $B_{L}^{opt}=(1-\sqrt{3})^{2}$ given by
\begin{equation} \label{optimalconstellation}
V_{\pm }=\left\{ \left( 
\begin{array}{c}
1 \\ 
0%
\end{array}%
\right) ,\frac{1}{\sqrt{2}}\left( 
\begin{array}{c}
1 \\ 
1%
\end{array}%
\right) ,\frac{1}{\sqrt{2}}\left( 
\begin{array}{c}
1 \\ 
\pm i%
\end{array}%
\right) ,\frac{1}{\sqrt{2}}\left( 
\begin{array}{c}
1 \\ 
\alpha (1\mp i)%
\end{array}%
\right) \right\} ,
\end{equation}
where $\alpha =(\sqrt{3}-1)/2$. Interestingly, the sets $V_{+}$ and $V_{-}$ both contain three MU bases plus one additional vector. 

\subsection{Dimension six}

According to Eq. (\ref{numberofequations}) the number of variables parameterising candidates for MU constellation increases quadratically with the number of undetermined vectors. Thus, a successful application of semidefinite programming in dimension two does not guarantee that a similar algorithm will terminate in dimension six. As for Gr\"obner bases, finite computational resources may limit the type of problems which can be solved. We will consider two MU constellations in $\mathbb{C}^6$ with small and large numbers of variables, respectively. 

The columns of the \emph{spectral matrix} \cite{tao04}
\begin{equation} \label{spectral}
S=\frac{1}{\sqrt{6}}\left( 
\begin{array}{cccccc}
1 & 1 & 1 & 1 & 1 & 1 \\ 
1 & 1 & \omega  & \omega  & \omega ^{2} & \omega ^{2} \\ 
1 & \omega  & 1 & \omega ^{2} & \omega ^{2} & \omega  \\ 
1 & \omega  & \omega ^{2} & 1 & \omega  & \omega ^{2} \\ 
1 & \omega ^{2} & \omega ^{2} & \omega  & 1 & \omega  \\ 
1 & \omega ^{2} & \omega  & \omega ^{2} & \omega  & 1%
\end{array}%
\right) \, ,
\end{equation}
define an orthonormal basis of $\mathbb{C}^6$ since $S$ is unitary. It has been shown that the pair $\{I,S\}$ cannot be extended to a triple of MU bases \cite{brierley+09} (the identity $I$ is associated with the standard basis of $\mathbb{C}^6$). Furthermore, there is no pair of orthogonal vectors $\{|u\rangle ,|v\rangle \}$ MU to the column vectors of $I$ and $S$. The two undetermined vectors $\{|u\rangle ,|v\rangle \}$ depend on 20 real variables which must satisfy 21 constraints given by $4^{\rm{th}}$ order polynomials. Due to the relatively small number of variables involved, one may expect that this property can be checked using semidefinite programming. Upon transforming the original problem, the equivalent SDP has been found to require $10,625$ decision variables satisfying $4,851$ linear constraints while the semidefinite inequalities involve matrices of dimension $231\times 231$---at the lowest level of relaxation, $r=2$. It took approximately 210 minutes and 5.4G of memory for the SDP to terminate, resulting in a global lower bound for the square of one of the polynomials defined by the MU conditions,
\begin{equation} \label{specialbound}
B_{L}(2)=2.28\times 10^{-8} \, .
\end{equation}
Since $B_{L}(2)$ is \emph{positive}, the SDP algorithm confirms the result of \cite{brierley+09}: no two orthonormal vectors can be found which are mutually unbiased to the column vectors of the pair $\{I,S\}$. At the next level of relaxation, $r=3$, the resulting SDP is already too large to terminate: there are $230,229$ variables while the constraints involve semidefinite matrices of dimension $1771\times 1771$. Consequently, we have not been able to improve the lower bound (\ref{specialbound}).

It should not come as a surprise our attempts to construct the SDP for the constellation $\{5^{3},1\}_{6}$ have not been successful, even at the lowest level of relaxation. Judging by the increase in the number of decision variables and the size of the semidefinite constraints seen in the previous examples, it is likely to be very large. Solving the resulting SDP would seem optimistic. However, it is possible that the resulting SDP has some specific structure which would allow one to use appropriate techniques.

\section{Summary and outlook}

We have reviewed two known computer-aided methods suitable to obtain \emph{rigorous} bounds on the number of MU bases existing in the space $\mathbb{C}^d$, based on a discretization of parameter space and on Gr\"obner bases. Both methods are currently too costly from a computational point of view to decide on the (non-) existence of more than three MU bases in dimension six. 

In addition, we have rephrased the problem in terms of semidefinite programming which allows one to use rigorous methods of \emph{global optimization}. This approach has been shown to reproduce two known results, the non-existence of four MU bases in dimension two, and the fact that the spectral matrix $S$ cannot be part of a complete set of seven mutually unbiased bases in dimension six.

Seemingly innocent questions about sets of mutually unbiased vectors in low-dimensional Hilbert spaces such as $\mathbb{C}^6$ continue to resist a variety of rigorous attempts of solution, confirming once more the dictum 
\cite{caves+96}
\begin{quote}
\emph{Hilbert space is a big place.}
\end{quote}

Interestingly, some MU constellations in dimension six are just within reach of \emph{numerical} methods \cite{brierley+08} which, however, only make their non-existence plausible. It remains to be seen whether a proof of the (non-) existence of a MU constellation such as $\{5^3,1\}_6$ will be found first by a computer-aided approach or by a purely analytic reasoning. It is safe to say though that obtaining a proof will be a matter of time only:  
the algorithmic methods presented here ascertain that we do \emph{not} deal with 
an \emph{undecidable} question when searching for complete sets of MU bases in 
composite dimensions.

\end{document}